\documentclass[conference,letter]{IEEEtran18}
\usepackage{xcolor}
\usepackage{url}
\usepackage{caption}
\usepackage{tcolorbox}
\usepackage{graphicx}
\usepackage{hyperref}
\usepackage{makecell}

\begin{document}
    \title{RASAECO: Requirements Analysis of Software for the AECO Industry}
    
    \author{
        \IEEEauthorblockN{Marko Ristin\IEEEauthorrefmark{1}, Dag Fjeld Edvardsen\IEEEauthorrefmark{2},
        Hans Wernher van de Venn\IEEEauthorrefmark{1}}
        \vspace{1ex}
        \IEEEauthorblockA{%
            \IEEEauthorrefmark{1}Zurich University of Applied Sciences (ZHAW), \{rist, vhns\}@zhaw.ch}
        \IEEEauthorblockA{%
            \IEEEauthorrefmark{2}Catenda AS, dag.fjeld.edvardsen@catenda.no}
    }

    \maketitle
    
    \begin{abstract}
    Digitalization is forging its path in the architecture, construction, engineering, operation (AECO) industry.
    This trend demands not only solutions for data governance but also sophisticated cyber-physical systems with a high variety of stakeholder background and 
    very complex requirements.
    Existing approaches to general requirements engineering ignore the context of the AECO industry.
    This makes it harder for the software engineers usually lacking the knowledge of the industry context to elicit, analyze and structure the requirements and to effectively communicate with AECO professionals.
    To live up to that task, we present an approach and a tool for collecting AECO-specific software requirements with the aim to foster reuse and leverage domain knowledge.
    We introduce a common scenario space, propose a novel choice of an ubiquitous language well-suited for this particular industry and develop a systematic way to refine the scenario ontologies based on the exploration of the scenario space.
    The viability of our approach is demonstrated on an ontology of 20 practical scenarios from a large project aiming to develop a digital twin of a construction site.
\end{abstract}

    \definecolor{airforceblue}{rgb}{0.36, 0.54, 0.66}
    \definecolor{asparagus}{rgb}{0.53, 0.66, 0.42}
    \definecolor{amaranth}{rgb}{0.9, 0.17, 0.31}
    
    \section{Introduction}

Projects, big and small alike, need to anticipate user expectations and scope
in form of requirements.
This is especially true for large software projects in the architecture-construction-engineering-operation (AECO) industry: they tend to be heavily invested in not only software complexity, but also complex business logic, industry practices, legal rules and a myriad of stakeholders with non-aligned interests and different technical backgrounds~\cite{systems_integration_in_architecture}.
This complexity makes the understanding between software engineers (programmers,
software architects \emph{etc.}) and non-software engineers (construction engineers, business people,
lawyers \emph{etc.}) paramount for project success.

General software projects have both the problem and the solution space open, \emph{i.e.}, the engineers developing the system need to capture the stakeholder requirements (the problem space) as well as come up with the system requirements (the solution space)~\cite{requirements_engineering_book}.
In contrast, and analogous to fields such as smart manufacturing~\cite{rami}, the requirements engineering of the software for AECO industry need to consider many procedures, best practices and regulations already established in the industry.
The stakeholder requirements and thus the problem space is given to a large extent, as opposed to general projects.
Consequently, the requirements engineering shifts to the solution space to identify system requirements in the specific context of the industry.

To that end, we tailor a scenario-based approach~\cite{glinz_scenarios} to requirements analysis suited for large AECO software projects addressing the following questions about the solution space:

\begin{itemize}
    \item What do the scenarios share \emph{in the context of AECO}?
	\item What \emph{language} should we use to make the scenarios more formal and readable by both requirements engineers \emph{and} AECO professionals?
    \item How can we refine the scenarios \emph{systematically}?
    \item How can we \emph{reuse} the results \emph{between} the projects?
\end{itemize}

For example, consider the case of risk management on a construction site.
There exist standardized procedures how risks should be identified, tracked and managed.
There are predefined actions that need to be undertaken in specific phases of the building lifecycle (\emph{e.g.}, risks that are considered during the planning phase and then tracked during the construction phase) and at different levels of abstraction (\emph{e.g.}, person-specific risks and site-specific risks, respectively).
The nomenclature and the data structures such as Building Information Modeling (BIM~\cite{IFC_ISO_Standard}) are widely used in the industry and even mandatory in many publicly-funded construction projects.

Mapping such procedures to a software system as ``black boxes'' ignoring the AECO context is usually tedious and wasteful.
The terms for both functional and non-functional requirements, and their grouping and organization need to be constantly reinvented from project to project.
This is particularly hard if the requirements engineer lacks the domain expertise, which can often be assumed to be the case in practice.

In contrast, we propose a method to analyze the requirements for software in the AECO industry.
We argue that it is much easier if the structure of the solution space is already preset and needs to be ``filled'' rather then re-invented each time.
We developed a conceptual framework to capture the scenarios in a structured manner with the  industry context in mind, including not only the domain of data governance but also cyber-physical systems. Our contribution is four-fold:

\begin{itemize}
    \item We introduce a novel \emph{scenario space} to highlight the common dimensions of the AECO procedures.
    \item Instead of common languages, such as UML~\cite{bim_governance_bpmn_uml}, we explore a novel choice and use Building Information Model (BIM) and its open standard Industry Foundation Classes (IFC~\cite{IFC_ISO_Standard}) as our \emph{main modelling language}.
    \item The relations between scenarios can be thought of as an ontology.
    We propose a novel systematic approach to refining the AECO scenario ontologies based on our predefined and uniform space.
    The wasteful random explorations are minimized and reusable parts outlive the individual projects.
    \item Finally, as appropriate tools are crucial for requirements analysis~\cite{re_in_aeronautics}, we provide \emph{a software tool} to analyze and visualize the resulting requirements~\cite{rasaeco_tool}.
\end{itemize}

We call our approach ``Requirements Analysis of Software for the AECO Industry'' or RASAECO, for short.

In Section~\ref{sec:related_work}, we summarize the prior art and motivate
our approach.
Sections~\ref{sec:project} and~\ref{sec:our_approach} describe the development environment of the approach, a large AECO software project, and explain the approach in detail accompanied by motivating examples, respectively.
section~\ref{sec:our_tool} explains how an open source tool developed in this project helps analyze the requirements in a scalable manner.
Section~\ref{sec:evaluation} describes how we applied the approach on gathering and analyzing the requirements for the project.
Section~\ref{sec:threats} investigates how well the approach generalizes to other projects in AECO industry.
We conclude the work in Section~\ref{sec:future_work} and examine the limits of our approach and future work.

    \section{Motivation and Related Work}\label{sec:related_work}

Requirements engineering is a well-established branch of software engineering and many other industries, and is also finding its ways into AECO~\cite{starchitects}.

As the rich body of literature suggests, domain-specific requirements engineering of software is necessary and beneficial~\cite{re_in_aeronautics}.
Yet there is a limited number of works about how to approach the requirements engineering for AECO-related software.
There exist collections of AECO use cases (e.g.,~\cite{buildingsmart_use_cases}), but they lack a uniform solution space.
Therefore the cases are narrow and disconnected, which does not promote clarity.

There are studies on the application of existing \emph{conventional} requirements engineering techniques on the data governance software for the AECO industry~\cite{arayici, bim_governance, shafiq, bim_user_requirements, re_ui_in_aeco, re_for_aeco_simulation_tools}.
Their narrow focus and the disregard of the specifics of the industry make them ill-suited for generalization to wider range of AECO systems (\emph{e.g.}, cyber-physical ones).

Process models based on the languages (such as IDM~\cite{idm_too_specific} and BPMN~\cite{bim_governance_bpmn_uml}) can be included in the requirements (\emph{e.g.}, if a full specification is required like in~\cite{bim_governance_bpmn_uml}), but their focus on processes is too elaborate for the general requirements analysis~\cite{idm_too_specific}.

Concrete scenarios for cyber-physical systems in AECO were explored in~\cite{cps_aeco, cps_aeco1}.
Our work does not present concrete scenarios, but a framework how to structure and investigate such scenarios, including the unstructured scenarios given in~\cite{cps_aeco, cps_aeco1}.

The general tools for specifying the requirements are abundant.
It has been widely recognized that a successful project needs key requirements~\cite{requirements_engineering_book} to support finding a feasible solution in a targeted manner, where structure~\cite{requirements_in_the_loop}, semantic annotations~\cite{frame_annotations}, requirements templates and patterns~\cite{requirements_reuse}, and ontologies~\cite{guitar_ontology} are all beneficial.
In this vein, we provide a tool for structuring \emph{AECO-specific} requirements annotating them with \emph{AECO-specific} semantic tags and refining their ontologies in a systematic \emph{AECO-specific} way.
To the best of our knowledge, it is a first of its kind in this particular industry.
In contrast to sophisticated annotation tools such as ReqPat~\cite{reqpat}, our annotations are light-weight and provide only guidance to the reader as modelling at the text level proved too high a barrier for AECO professionals (see Section~\ref{subsec:ubiquituous-language}).

Visualization of requirements is important~\cite{visualization_review}.
Our work visualizes the requirements in spaces similar to the existing approaches in requirements engineering (\emph{e.g.}~\cite{shapeRE}).

There is also a fruitful parallel research in context of smart manufacturing to specify a solution space (termed ``reference architecture'') to structure the discussions, and organize standards and implementations (RAMI4.0~\cite{rami} and many others).
Our approach is strongly influenced by these trends and builds a scenario space for AECO following this pioneering work.

    \section{The BIMprove Project} \label{sec:project}

The current approach was incrementally developed during the BIMprove Project~\cite{bimprove}, an initiative funded by the European Union's Horizon 2020 Research and Innovation Program with focus on building a low-carbon, climate-resilient future.
The project started in September 2020 and aims to develop a dynamic digital system for the construction industry, increase productivity, cut costs and improve the working conditions.
The system should deal both with static (``as-planned'') as well as dynamic real-time data (``as-built'' or ``as-observed'').
The participants include 12 partners (2 universities, 3 research centers, 1 non-profit organization and 6 industrial partners).
The project will take 3 years with a total budget of 5.6 Mio. euros.
The team involves more than 40 people including 6 software engineers, 15 experts in the AECO domain, 13 experts in robotics and mechanical engineering,  9 user experience experts, 2 experts in standardization \emph{etc}.
The authors of this work also participate in the project: two authors with a decade-long experience in the AECO industry and one author with a decade-long experience in software engineering.

The project spans a large and variate scope: deviation between ``as-built'' \emph{versus} ``as-planned'' in many different settings (including augmented and virtual reality), enforcing cleanliness on the site, planning, tracking, and preventing risks, giving guidance on the site (\emph{e.g.}, for deliveries or work), \emph{etc.}
The project is expected to result in an experimental system ready for further development and deployment.

    \section{Our Approach: RASAECO}\label{sec:our_approach}

\subsection{System Scenarios} \label{subsec:system_scenarios}
We propose to analyze requirements in software projects within the AECO industry based on \textbf{scenarios}.
Scenarios are a popular tool in requirements engineering~\cite{glinz_scenarios} in which the requirements are gathered through reasoning about typical (real or imagined) user activities and system-user interactions.
In our context, a scenario is an abstract formal or semi-formal descriptions of an AECO process at the level of a system or a model.
It is not a narrative, but a structured document written by requirements engineers giving a process description from the perspective of different protagonists in the construction and operation process of a building.

As our domain entails cyber-physical elements, we explicitly shift the focus from the \emph{user} to the \emph{system} as recognized by~\cite{cps_aeco}.
In such systems the real physical world increasingly interacts with its virtual representation, making the end-user interactions equally relevant as interactions between the different system modules (and in advanced stages of a project, inter-system interactions).
Similar to ``system stories'' in~\cite{dronology}, we used the explicit term \textbf{system scenarios} whenever there was potential for misunderstandings
\footnote{%
The term ``use case'' or ``system use case'' would have been perhaps more precise and more in line with the existing literature~\cite{requirements_engineering_book}.
However, we discarded it since it sowed too much confusion with the AECO professionals in our case, where the term ``scenario'' or ``system scenario'' was much better understood and used.
This matches the colloquial usage of the term that was also employed in some works, \emph{e.g.}~\cite{cps_aeco}.}.

\subsection{Scenario Space}\label{subsec:scenario-space}

Requirements engineers, especially the novices without the AECO experience, can be at loss how to compile scenarios~\cite{novices_in_requirements}.
The boundless scenario space provides little reuse between and within projects~\cite{requirements_reuse} and the requirements engineers need to re-define it for each problem at hand.
Furthermore, an open scenario space also puts the burden on the reader as there might be no inter-project or even inter-scenario consistency thus forcing the reader to repeatedly infer it.

We therefore propose to set up an AECO-specific \textbf{scenario space} as a representation of any action taking place in the dimensions of the AECO industry.
This should make our scenarios uniform and easier to both write and read by dissecting the requirements along the predefined dimensions.
We propose the following three dimensions:
\begin{enumerate}
    \item Aspects,
    \item Phases in the building lifecycle, and
    \item Hierarchy levels of detail.
\end{enumerate}

The scenario space is defined in such a way that any AECO process can be mapped within these three dimensions.
A system fulfilling the complete scenario space is thus capable of processing upcoming real-world events (correction of the building structure, rescheduling, security measures \emph{etc.}) within the system.
Consequently, the system design and its implementation are easier to model and can be explicitly traced back to the scenario space.
Similar to software modules in a computer program, an individual scenario occupies only a portion of the scenario space by design, and we explicitly do not strive for scenarios which cover the whole space, as this would be both impractical to write and overwhelming to read.
Instead, we disassemble the more complex scenarios into less complex ones, allow concrete scenarios to specify the details of the more abstract ones, and relate the scenarios among themselves  (see Section~\ref{subsec:ontology}).
For example, a complex AECO procedure can be split into a series of steps, where each scenario represents a single step, and the bird-view of the procedure is encapsulated in a bundling scenario.
The final 3D puzzle of all the scenarios is expected to fully populate the relevant scenario subspace of the project.

\textbf{Aspects} address a concern of a multi-layered scenario.
Unlike classical requirements engineering, which usually groups requirements into functional, non-functional and technical~\cite{bim_governance}, and do not convey any domain-specific meaning, our aspects are specially tailored for the AECO industry, influenced by ``higher'' (5D, 6D \emph{etc.}) BIM dimensions~\cite{bim_dimensionality}.
We identify the following aspects:
\begin{itemize}
    \item \emph{As-planned}: models of plans and planned activities (both in space and over time),
    \item \emph{As-observed}: models of observations (both in space and over time),
    \item \emph{Divergence}: difference between the as-planned and the observed data,
    \item \emph{Scheduling}: logical order of activities and how it affects overall schedule,
    \item \emph{Cost}: financial background of the scenario,
    \item \emph{Safety}: safety regulations of workers and other actors (legal and non-legal) concerning the scenario and how the system should accommodate them,
    \item \emph{Analytics}: expected analytics, both technical (\emph{e.g.}, pipe routings, static and dynamic re-calculations) and business-related (\emph{e.g.}, key performance indicators).
\end{itemize}

Fig.~\ref{tab:aspects} gives reduced examples of the aspects of a real-world AECO scenario from our project (see Section~\ref{sec:project}).

\begin{table}
    \begin{center}
        \begin{tabular}{ll}
            \multicolumn{2}{l}{a) Truck Guidance} \\ [0.5ex]
            \multicolumn{2}{l}{``External truck drivers arrive at the construction site to deliver cargo.''} \\
            As-planned  & The deliveries are specified as tasks.                    \\
            As-observed & The driver's device tracks the GPS location.              \\
            Divergence  & The device guides the driver to the delivery location.    \\
            Scheduling  & The unmet deadlines are pushed as topics.                 \\
            Analytics   & The statistics of delivery delays are reported. \\ [2ex]

            \multicolumn{2}{l}{b) Risk Management} \\ [0.5ex]
            \multicolumn{2}{l}{``Different risks are planned and tracked on the site.''} \\
            As-planned  & The risk manager manages the risks.                       \\
            As-observed & The risk manager orders focus spots for the recording.    \\
            Divergence  & The preventive resource visually inspects the recordings. \\
            Scheduling  & The risk manager inspects the risk linked to the tasks.   \\
            Safety      & The risk manager marks the dangerous tasks.               \\
            Analytics   & The system tracks the escalated risks per category.  \\ [2ex]

            \multicolumn{2}{l}{c) Cost Tracking} \\ [0.5ex]
            \multicolumn{2}{l}{``The planned costs are tracked against the expenditures.''} \\
            Cost        & Planned costs and expenditures are related to tasks.      \\
            Analytics   & Over-budget tasks are reported.                           \\
        \end{tabular}
    \end{center}
    \captionof{table}{%
    Summarized aspects of three scenarios from~\cite{bimprove_scenarios}.
    The aspects provide support for the requirements engineer during writing as well as for AECO experts during reading and reviewing, in particular given their familiarity with them through the ``higher'' BIM dimensions~\cite{bim_dimensionality}.
    In contrast, conventional requirements analysis would have grouped them in ways unintuitive to AECO experts such as functional, non-functional and system requirements, while the requirements engineers without the industrial experience would probably miss some of the important AECO-specific aspects (see also Section~\ref{sec:evaluation}).
    }
    \label{tab:aspects}
\end{table}

A large project usually spans multiple \textbf{phases of the building lifecycle}.
Consequently, the requirements in a scenario follow the lifecycle, but often cannot be structured in a linear manner with the respect to time as other groupings might make more sense (\emph{e.g.}, grouping requirements by functionality).
The reader should thus always be put in context with clear references to phases in case of ambiguities.

How the phases should be split depends on many factors such as organizational and contractual criteria.
We modeled the phases based on business practices of our industrial partners in the project~\cite{bimprove}, but alternatives such as RIBA~\cite{riba} are equally plausible.
The phases thus depend on a concrete project, company, national and international environment.
We propose the following ones:

\begin{itemize}
    \item Planning,
    \item Construction,
    \item Operation,
    \item Renovation, and
    \item Demolition.
\end{itemize}

The relevant parts of the scenario document should be appropriately marked with the corresponding phase.
Figure~\ref{fig:phase_marking} shows an example of such a marked requirement.

\begin{figure}
    \begin{tcolorbox}[boxrule=0.5pt]
    {\color{asparagus}%
    The risk manager inputs the initial set of risks already known during the planning.\textsuperscript{planning}}

    \vspace{2ex}

    {\color{asparagus}%
    Both the risk manager and the preventive resource can insert new and change existing risks accordingly.\textsuperscript{construction}}
    \end{tcolorbox}

    \captionof{figure}{%
    Example of a requirement from Table~\ref{tab:aspects}~b) ``Risk Management'' marked with phases in the building lifecycle.
    In this example, it was initially not clear that the second sentence referred to the ``construction'' instead of the ``planning'' phase.
    The conventional approach to requirements analysis would ignore the phases or refer to them implicitly.
    Our approach provides a standardized way of how to consider the phases in the building life cycle, thus making it easier both for writers and readers to deal with them explicitly.}
    \label{fig:phase_marking}
\end{figure}

Complex scenarios require the analysis of multiple \textbf{hierarchy levels of detail} in parallel~\cite{requirements_zoom}, from a device to the network of companies, from component-level to system-level, with multiple levels interacting with each other.
To that end, we propose the following hierarchy inspired by RAMI4.0~\cite{rami} to capture and situate requirements in multiple abstraction levels:

\begin{itemize}
    \item Device or Person (\emph{e.g.}, a shovel, or an electrician),
    \item Machine or Crew (\emph{e.g.}, excavator, or a team of electricians),
    \item Site unit (\emph{e.g.}, a certain zone on the construction site),
    \item Site,
    \item Site office,
    \item Company, and
    \item Network of companies.
\end{itemize}

The hierarchy follows the organization of the AECO industry and how it is decomposed into units: a device is a hand tool or a part of a bigger system (\emph{e.g.}, a smartphone as a part of the communication system), a machine is a more complex system and can contain different devices (\emph{e.g.}, an excavator containing a shovel, hydraulic cylinders \emph{etc.}), a site unit is responsible for machines and teams \emph{etc.}
The levels should therefore allow for more straightforward mapping from the AECO procedures to software requirements.\footnote{%
We originally entertained a thought to organize the levels in a \emph{tree} instead of a \emph{list of levels}, but the trees we explored were too confusing and cumbersome to use in practice, so we decided to use the simple hierarchy akin to the ``height'' of the viewer (ground-view \emph{versus} bird-view).
}
The levels of detail in the hierarchy are marked in the document analogous to phases. 
Figure~\ref{fig:level_marking} presents an example marking.

\begin{figure}
    \begin{tcolorbox}[boxrule=0.5pt]
    {\color{amaranth}%
    The truck driver's device tracks the GPS location, but does not send it to the system.\textsuperscript{device}}

    \vspace{2ex}

    {\color{amaranth}%
    The location is only used to navigate the truck.\textsuperscript{machine}}
    \end{tcolorbox}

    \captionof{figure}{Example of a requirement from Table~\ref{tab:aspects}a) ``Truck Guidance'' marked to highlight the level of detail.
    The marking enhances the reading by putting the extra accent.
    The reader is made aware that the location is only relevant for the machine (truck), but not the system.
    The predefined abstraction levels allow for consistent markings throughout the project while arbitrary markings usually employed in conventional requirements engineering would differ from scenario to scenario.
    This makes it easier for the writer using our approach to label the text as no custom levels need to be invented, but can be readily picked off-the-shelf.
    At the same time, the consistency of the levels throughout the project makes the reading also easier.
    }
    \label{fig:level_marking}
\end{figure}

\begin{figure}
    \includegraphics[scale=0.8]{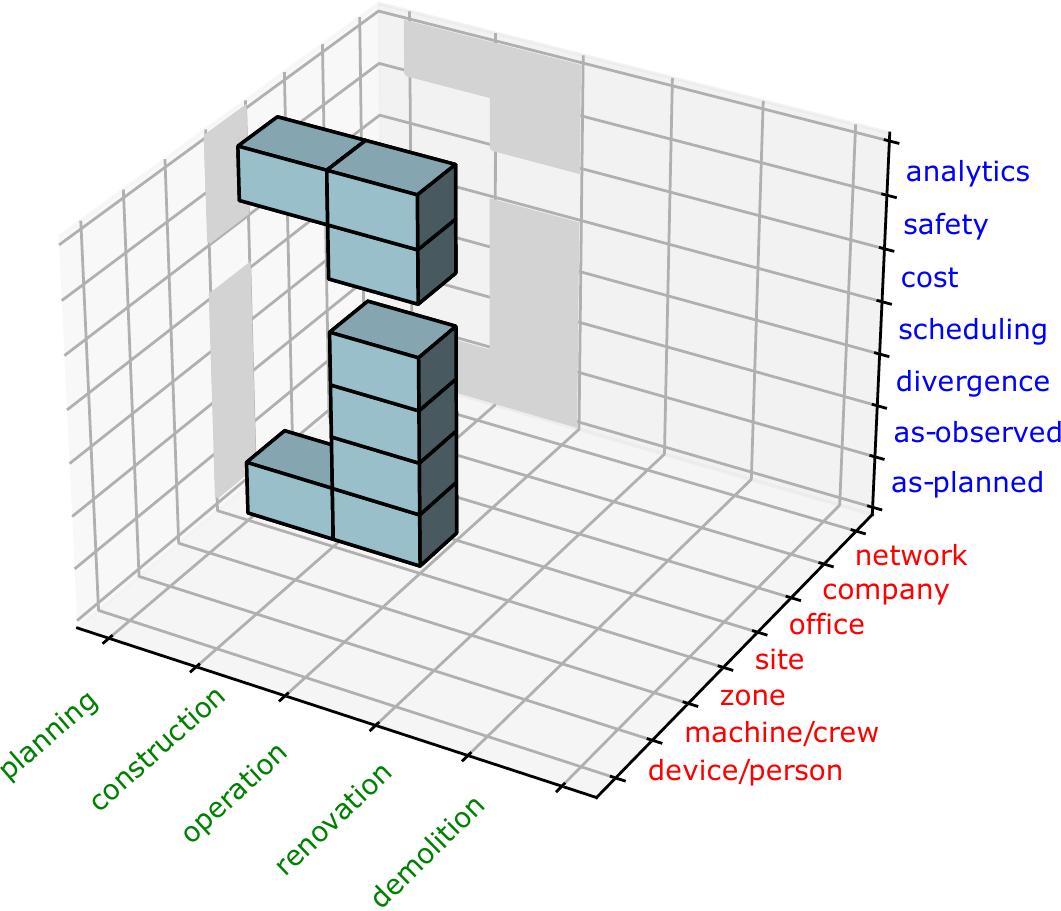}

    \captionof{figure}{Visualization of the scenario from Table~\ref{tab:aspects}b) ``Risk Management'' as a volumetric plot.
    Such a visualization is practical when scenarios are summarized as images.
    As we can see from the plot, the scenario ``Risk Management'' is focused on the site risks (such as injuries due to missing safety equipment and unfortunate site configurations or potential for fires due to spatio-temporal proximity of dangerous tasks).
    This scenario is not concerned about the individual site-independent risks related to device/person, machine/crew or zone, nor does it cover the risks that can affect the company or the network (\emph{e.g.}, the risks concerning cost and scheduling).
    These risks are covered by a different scenario and were left out here for reasons of clarity.}
    \label{fig:space}
\end{figure}

Employing three dimensions allows for a nice metaphorical visualization of the scenario space as a volumetric (\emph{i.e.}, individual scenarios are represented as a set of cubes  in that 3D space).
Figure~\ref{fig:space} demonstrates such a visualization of an example scenario.
This visualization is especially practical when multiple scenarios are displayed, \emph{e.g.}, in a list accompanied by thumbnail icons, or showing how complex scenarios are assembled together to form a more comprehensive 3D ``puzzle''.

\subsection{Ubiquitous Language} \label{subsec:ubiquituous-language}

Domain knowledge is important for capturing requirements~\cite{domain_knowledge_important}.
Representing the domain with a ubiquitous language is an essential element for uncovering ambiguities and misunderstandings between software engineers and domain experts~\cite{domain_drive_design}.
In our setting, it is essential to translate the language of the construction industry into the language of software engineers if a project is to succeed.
Unified Modelling Language (UML), a software engineer’s choice, is often inappropriate for communication with AECO professionals unfamiliar with that language~\cite{arayici, idm_too_specific, uml_too_heavy_weight}.

Instead, we propose to use the \textbf{Building Information Model (BIM)}~\cite{IFC_ISO_Standard} as a formal language for capturing requirements and describing the system entities, relationships and behavior in our scenarios. 
Thankfully, the AECO industry enjoys the luxury of an increasing number of professionals getting versed in it world-wide (from Peru~\cite{bim_in_peru} to Malaysia~\cite{bim_in_malaysia}).
Using BIM as a language for \emph{software requirements} is novel as it is still predominantly regarded rather as an exchange or storage format, and to the best of our knowledge, it has not been used as ubiquitous language so far.

Though it is important to avoid ambiguities, it is equally important to avoid formalisms and additional notation whenever possible to lower the entry barrier and \textbf{"keep it simple"}~\cite{listens_learned}.
Visual notation should be used sparingly and as comprehensible as possible~\cite{comprehensible_visual_notation}.
In our experience (see Section~\ref{sec:evaluation}), the AECO professionals lacking computer science background were unfamiliar with formalisms such as pseudo-code and structured text, and were rather put off and blocked to start the analysis too formally.
Moreover, structuring the problem upfront can be detrimental as formalisms stifle creativity and impede the transfer of domain knowledge, while sense-making can nourish it~\cite{sensemaking}.
We recommend starting with an abstract informal scenario and refining it only if utmost necessary.

In our system scenarios~\cite{bimprove_scenarios}, however, we did not find a single appropriate spot where requirements had to be formalized.
What really worked well to clear up the ambiguity was reliance on BIM concepts such as identifying appropriate IFC classes and relationships.
This helped us better discuss what data is relevant (and what can be left out), and often highlighted the data flow through our system.
Figure~\ref{fig:bim} shows a requirement where BIM is used for more precise definitions.

\begin{figure}
    \begin{tcolorbox}[boxrule=0.5pt]
        \textbf{Performance history} is defined as an instance of \texttt{IfcPerfromanceHistory} and
        lives in the model \texttt{bim\_extended}.

        \vspace{2ex}

        \textbf{Cost} is an instance of \texttt{IfcCostItem} living in the model \texttt{bim\_extended}.
        It can be linked to \emph{tasks} through GUIDs and \texttt{IfcRelAssignsToControl} (where
        the \emph{task} is the related object).

        \vspace{2ex}

        \textbf{Expenditure} is an \texttt{IfcCostItem} living in the model \texttt{bim\_extended}
        together with its relations.

        To distinguish it from estimated \emph{costs}, expenditures are explicitly linked to a performance history through \texttt{IfcRelAssignsToControl} (where the performance history is the control and the expenditure the related object).
    \end{tcolorbox}

    \captionof{figure}{%
    Example of a requirement from Table~\ref{tab:aspects}c) ``Cost Tracking''
    where Building Information Modelling (BIM) helped us disambiguate 
    the concepts and precisely express the relations in a way 
    familiar to AECO professionals.
    Conventional approaches to requirements engineering use a general language 
    such as UML which forces the invention of custom concepts and relations, 
    and incurs the overhead since AECO experts not only need to learn 
    a new language, but also need to understand these custom concepts and 
    relations.
    In this example, costs and expenditures can be thought 
    in many different ways, but framing them as \texttt{IfcCostItem}\'s and 
    their relations to other entities as \texttt{IfcRelAssignsToControl} 
    make the reader immediately aware of intended hierarchies such as those 
    stemming from \texttt{IfcControl}.
    }
    \label{fig:bim}
\end{figure}

In the future, we will develop a tool for the graphic formulation of scenarios to support less tech-inclined users with describing the scenarios using BIM, and composing simpler scenarios into more complex ones, to further minimize the need for formalisms.

\subsection{Scenario Ontology} \label{subsec:ontology}

Scenarios span a spectrum from very simple action sequences (\emph{e.g.}, sending warning messages) to extremely complex AECO processes (demolition and new construction of building structures, including rescheduling
\emph{etc.}).
As their numbers increase, we need to organize and relate them to each other to obtain a holistic view of the requirements.
An ontology is a common strategy to organize requirements (\emph{e.g.}~\cite{glinz_statecharts, guitar_ontology}).

To provide the reader with a big picture, we organize and present the relations between the scenarios in a \emph{scenario ontology} following different vectors.
We adopt the existing approaches: the scenarios can be related as steps of a sequence or a state-chart~\cite{glinz_statecharts}, or as abstract-concrete relations similar to sub-typing~\cite{requirements_zoom}.

While the conventional approaches model the ontologies well and can be readily integrated for the AECO scenarios, our approach really shines when the ontology is defined in terms of the scenario space, where complex scenarios are \textbf{dissected by dimensions} of the scenario space.
Instead of arbitrary criteria, we can systematically refine the ontology.
We start from an initial set of scenarios and explore the space along its dimensions.
We focus either on the individual aspects or phases in the building lifecycle of a scenario, or ``zoom'' in and out along the hierarchy levels of abstraction.
This is akin to graphical exploration of~\cite{requirements_zoom}, but in a coherent manner specifically tailored to the context of the AECO industry (see also Section~\ref{subsec:effect_of_scenario_space_on_refinement}).

\begin{figure}
    \begin{center}
        \begin{tabular}{cc}
            a) & \includegraphics[scale=0.55]{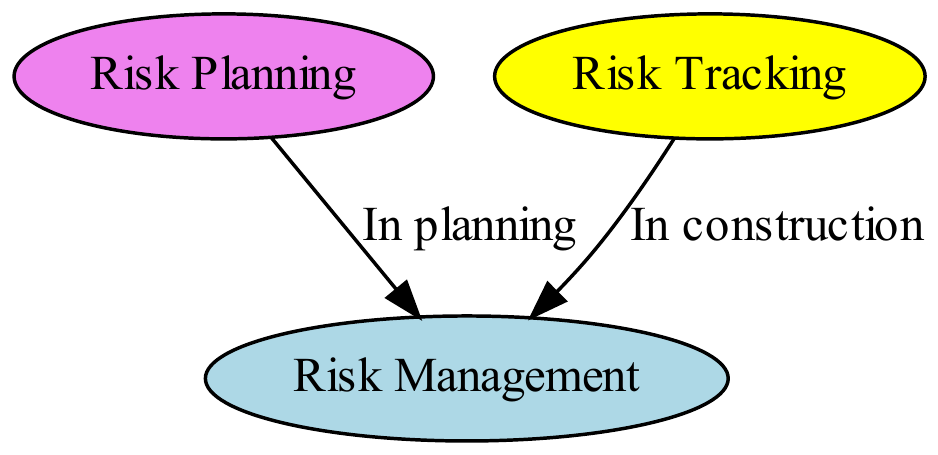} \\[1ex]
            b) & \includegraphics[scale=0.7]{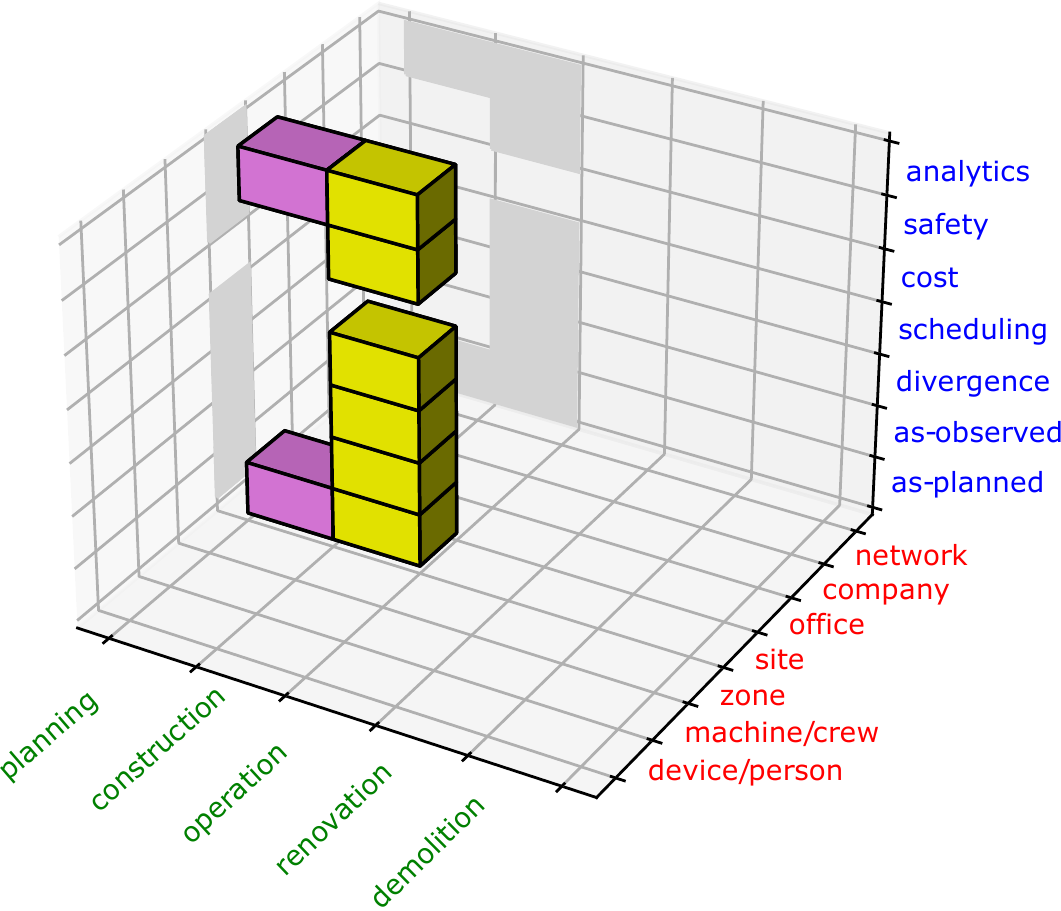}
        \end{tabular}
    \end{center}

    \captionof{figure}{%
    \textbf{a)} The subset of the ontology showing how the scenario ``Risk Management'' from Table~\ref{tab:aspects}b) is decomposed into two scenarios, \emph{Risk Planning} (violet) and \emph{Risk Tracking} (yellow) along the phases ``planning'' and ``construction''.
    The arrows represent the relations.
    \textbf{b)} The decomposition of the scenario \emph{Risk Planning} in the scenario space (see Section~\ref{subsec:scenario-space}).
    The ontology gives the reader a first impression of the decomposition while the volumetric is an additional comprehension aid.
    The more focused scenarios, ``Risk Planning'' and ``Risk Tracking'', outlay the details of risk management concerning the respective phases.
    For example, the focus of ``Risk Planning'' is on inserting and defining risks, while ``Risk Tracking'' covers how the planned risks are observed and, if necessary, escalated.
    As we can immediately see in the volumetric, ``Risk Planning'' does not enforce safety measures, while ``Risk Tracking'' includes them (by the aspect ``safety'').
    A standardized scenario space helps the writer to think in possible refinements and decompositions in a systematic way, and aids the comprehension of the reader at the same time.
    }
    \label{fig:split-by-phase}
\end{figure}

Not all possible scenarios need to be defined in advance as the completeness of requirements is rarely feasible in practice~\cite{glinz_scenarios}.
Additional scenarios are added and embedded into the scenario space at a later point in time, as the discussion with the domain experts and other stakeholders progresses.
This makes the scenario ontology a malleable structure, expected to evolve through the lifetime of the project in a systematic way.
Depending on the adoption within a company and a wider community, certain parts of the ontology are shared across multiple projects.
The most distilled parts are finally standardized to help requirements reuse~\cite{requirements_reuse}.
Thanks to a standardized scenario space, it is possible to identify and situate such reusable parts, and consequently allows for consistent reading.
Figure~\ref{fig:split-by-phase} gives an example of our refinement method and shows how a scenario is decomposed according to the phases of the building lifecycle.

\subsection{Document Structure} \label{subsec:document_structure}

While we do not consider the document structure to be crucial and believe that an alternative structure works equally well, we give a starting point for the practitioners here.
Our main reference was~\cite{lightweight_requirements_assesments}, in particular their document analysis.
The scenario document is generally structured by the aspects where each aspect is usually represented by a subsection.
We note that not all aspects need to be represented and irrelevant aspects can be omitted when appropriate.

\begin{enumerate}
    \item We start with a ``Summary'' section describing the scenario in succinct language.
    \item The section ``Relations to Other Scenarios'' specifies how the scenario relates to other scenarios.
    \item The section ``Models'' defines a mental map of the data, described in abstract terms.
    It sometimes includes relevant types, sources and the storage medium.
    \item Third section, ``Definitions'', defines the entities as well as their relationships.
    \item The fourth section, ``Scenario'', elaborates the scenario with each aspect making a separate subsection.
    Not all aspects need to be represented and irrelevant aspects are omitted when appropriate.
    \item The section ``Test cases'' describes how the implementation of the scenario is tested in practice.
    \item The last section, ``Acceptance Criteria'', indicates the non-functional requirements such as efficiency constraints, expected magnitude of the data \emph{etc.}
\end{enumerate}

The document is appended by an index to the parts of the document marked with different phases in the building lifecycle and hierarchy levels of detail.
We published the scenario documents in~\cite{bimprove_scenarios} and their source in~\cite{bimprove_scenarios_src}.

\section{Our Tool}\label{sec:our_tool}

Writing structured requirements is difficult and a specialized tool can provide some rails~\cite{quality_user_stories}.
To that end, we implemented a prototype command-line application to help us analyze requirements in a scalable manner.

We chose markdown as the document format for its ease of editing in a text editor and its multi-media support~\cite{markdown}.
The pseudo-code was marked with code blocks, while we used custom markups for markings (phases and hierarchy levels), references to definitions of models, entities, queries, commands and events.
The references between the scenarios in the ontology were captured in a \texttt{<rasaeco-meta>} element with a property "nature" along with other meta information such as the volumetric in the scenario space.
Since the source of the scenario document is in plain text, it can be readily fed into a version control system (such as Git~\cite{pro_git}).
We provide the source code of the scenario ontology tailored for our project~\cite{bimprove_scenarios} at~\cite{bimprove_scenarios_src}.

\begin{figure}
    \begin{center}
        \includegraphics[scale=0.55]{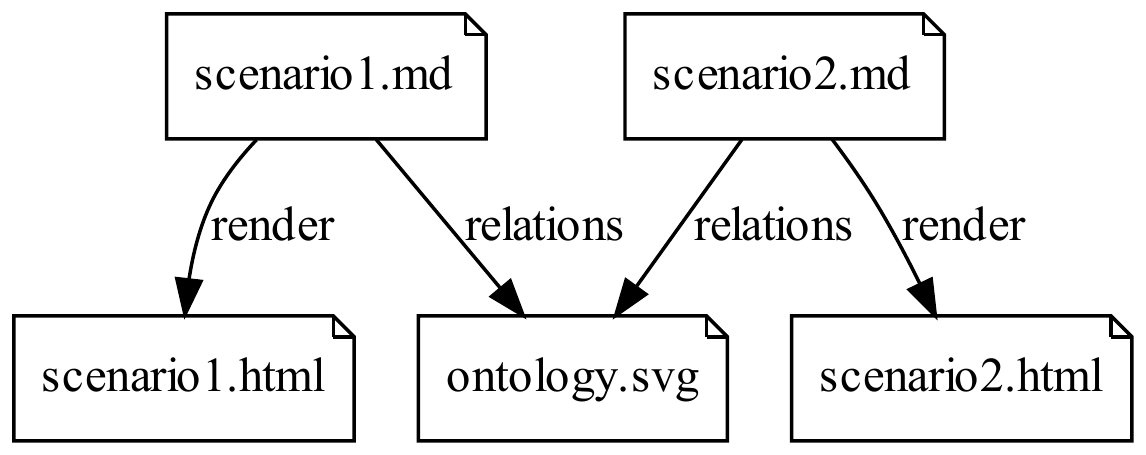}
    \end{center}

    \captionof{figure}{Overview of the rendering process of our tool.
    The markdown files (\texttt{scenario1.md} and \texttt{scenario2.md}) are rendered to HTML (\texttt{scenario1.html} and \texttt{scenario2.html}).
    The scenario ontology (\texttt{ontology.svg}) is rendered as a graph based on the relations defined in the meta-data header in these two markdown files.}
    \label{fig:processing}
\end{figure}

The tool converts the individual source documents (in markdown) to browser-readable HTML documents.
The set of all source documents is analyzed to reconstruct the ontology which is then also rendered as a graph (in HTML/SVG).
Figure~\ref{fig:processing} outlines the processing of the data from the source to the artifacts.

The tool was implemented in Python 3.8 and released under MIT license on \href{https://pypi.org}{pypi.org}.
The standalone binaries are available for 64-bit Windows on the web site~\cite{rasaeco_tool}.

    \section{An Experience Report on Applying RASAECO} \label{sec:evaluation}

In this paper we proposed a novel approach and a tool for analyzing and refining system scenarios in the AECO industry. 
Thus far, we focused our efforts on developing both the concepts and the tool to make them ready for practical use.
Due to limited resources, we could not evaluate our approach on a set of different projects and recognize that more systematic field studies and varied use cases are needed to assess RASAECO in a more rigorous manner.
Nevertheless, the initial results look promising.
Inspired by the experience report from~\cite{dronology}, we qualitatively assess how our approach helped us analyze the requirements for a large software project in AECO (see Section~\ref{sec:project}):
\begin{itemize}
    \item How did the proposed scenario space help with analyzing and structuring the scenarios?
	\item How did BIM as ubiquitous language affect the requirements analysis?
    \item How did our approach guide us with refining the scenarios (and how much effort could we re-use)?
\end{itemize}

\subsection{The Requirements Analysis} \label{subsec:requirements_analysis}

Here we illustrate our requirements analysis so that other projects can be related to its volume, scope and characteristics.

The requirements elicitation phase took the first 6 months of the project (including brainstorming sessions \emph{etc.}).
We had trouble recognizing the requirements in the beginning.
We started with a conventional approach, collecting the  requirements and organizing them by functional/non-functional/system requirements.
However, we faced multiple problems.
The requirements were collected at different levels of abstraction, but we lacked a uniform  notion of the levels and so had difficulties communicating within the team.
The requirements were laborious to group and categorize due to the wide scope of the project, and we had a lot of initial dissonance on categories.
Each contributor had their own notes, but the common grounding was difficult without a shared terminology.
Finally, each contributor had her own scenario dimensions in mind which provoked further misunderstandings.

Faced with these challenges, we developed our approach to address them in a systematic way.
The requirements analysis which we examine in this work was performed in the course of 3 weeks over 20 sessions at the end of this 6-months project phase, where each session took 2-2.5 hours.
The sessions were guided by a software engineer (an author of the study, with more than 10 years experience in software engineering, but no prior experience in AECO or related fields), where he interviewed 4-5 stakeholders relevant for the respective system scenario.
Each session covered a specific topic (\emph{e.g.}, ``Truck Guidance'' summarized in Table~\ref{tab:aspects}a), and what took place was something between an interview and a conversation.
As the sessions progressed, the previous scenarios were refined and refactored in addition to creating new ones.
The final scenarios comprised the whole scope of our system.

The analysis finally resulted in 20 system scenarios.
Most scenarios were rather short (8 scenarios with $<500$ words).
There were medium-long scenarios (7 with up to 1000 words) and a few long ones (5 with more than 1000 words).
Coincidentally, the long scenarios ($>1000$ words) encompassed the core modules of the system (such as  ``Digital Reconstruction'', ``UXV Recording'' and ``Virtual Inspection'') with fewer AECO-specific elements.
In contrast, AECO-specific scenarios, such as ``Risk Management'', could be further refined by our approach into smaller sub-scenarios, additionally resulting in a smaller word count  (all less than 1000 words).

The final ontology graph was not highly connected and remained manageable.
With respect to incoming edges, many scenarios had no incoming edge (9 scenarios).
There were a few scenarios with one or two incoming edges (5 and 4, respectively).
There was only one scenario with 3 and one with 4 incoming edges, respectively.
The outgoing edges were very few per scenario: some with zero edges (3 scenarios), most had a single outgoing edge (15), and only two scenarios had two and three outgoing edges, respectively.
Notably, the scenarios with many incoming edges included a core module (``Topic Management'', serving as a communication bus) and a complex scenario refined in multiple sub-scenarios (``Risk Management'').
The out-degree of the scenarios with more than one outgoing edge (``Digital Reconstruction'' and ``Risk Tracking'') reflects their data flows (pushing the information to multiple downstream scenarios). 

\begin{figure}
    \begin{center}
        \includegraphics[scale=1]{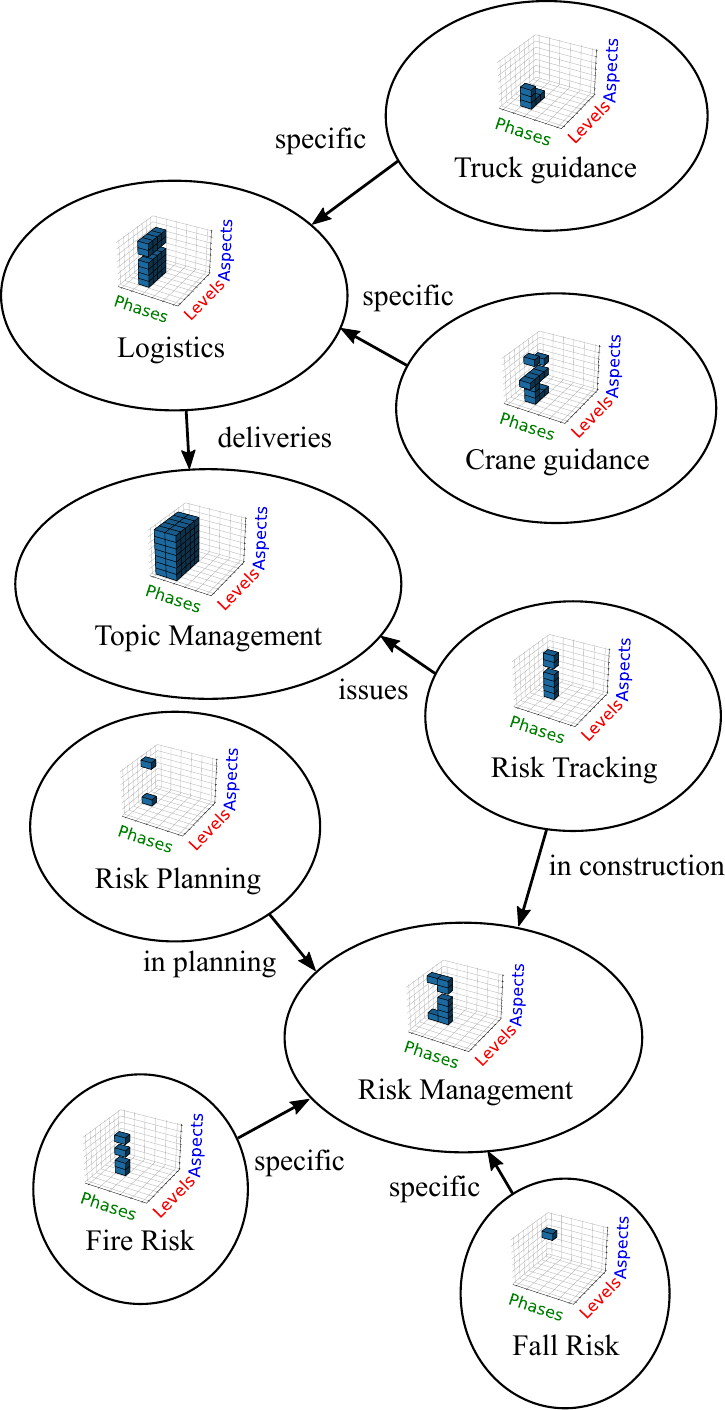}
    \end{center}

    \captionof{figure}{%
    The sub-graph of our ontology.
    The full ontology is available at~\cite{bimprove_scenarios}.
    This sub-graph gives a glimpse of the scope and complexity of our scenarios to be compared with similar ontologies.
    The volumetric plots are iconified intentionally and are not meant to be read in detail but rather to give a vague idea about the concerns of a scenario and its focus.
    A larger volumetric plot, such as one given in Figure~\ref{fig:space} is included in the text of the scenario.
    }
    \label{fig:sub-ontology}
\end{figure}

Constricted by the scope of the project, the volume of the scenarios is concentrated in the planning and construction phases (only 1 about planning, 14 about construction and 5 about both phases).
The abstraction levels reached ``office'' and spanned all aspects.
A sub-graph of our ontology in Figure~\ref{fig:sub-ontology} gives a glimpse of the scope and complexity of our ontology to be compared with other projects.
We discuss the generalizability to other AECO projects in Section~\ref{subsec:generalizability}.

\subsection{Effects of the Scenario Space on Analysis} \label{subsec:effect_of_scenario_space_on_analysis}

Here we observe the effects that the scenario space had on writing the individual scenarios.

\textbf{Degree of participation}. 
Our approach forced a certain structure on the authors, so that they had to at least be familiar with the relevant terms such as dimensions of the scenario space such as aspects, phases and levels.
Somewhat unexpectedly, thinking and writing in such structures proved to be too much of a barrier for AECO experts, so a software professional had to take over that task.
The resulting text was readable and something they could comment on and finally approve of, but could not contribute to freely.

\textbf{Discovering AECO-specific complexities}.
The relatively rigid structure forced on us an early positive confrontation between what is easy/difficult to implement and what is of low/high value from an AECO perspective as the structure prevented us to ignore the different dimensions of the solution space.
We experienced that our approach substantially framed our thinking in the terms of the solution space.
In particular, the tension between as-planned, as-observed and their divergence was very useful as it mapped well to many of the cyber-physical problems (key to 7 out of 20 scenarios) and forced us to define the sources of the data at an early stage.
Notably, the risk-related scenarios (``Risk Tracking'', ``Fire Risk'' and ``Fall Risk'') and delivery scenarios (``Truck Guidance'' and ``Crane Guidance'') could easily be mapped to this planned-observed schema.

Additionally, the distinction between cost, scheduling, analytics and safety helped us to distinguish the nuances and to group the requirements.
For example, while working on the scenarios related to logistics, we struggled with the opaque two-fold nature of the concept ``missed deadlines''.
Distinguishing explicitly between scheduling and analytics finally identified the ``missed deadlines'' as an analytics indicator and a separate trigger for re-scheduling.
In a similar vein, the phases allowed us to think about the temporal sequence of the logistics workflow: entrances and exits are defined during planning (and rarely updated later), while delivery locations change frequently during the construction phase.

For us this indicates that our method is well-suited to drilling down into important details through deep discussions at the core of the structured scenarios.
As the structure forced us to quickly identify and clarify important \emph{AECO} unknowns, unclear parts of the AECO procedure were revealed.
However, the experiences of using the method in additional projects is still needed to obtain better validation.

\textbf{Markings as a disambiguation instrument}.
We used semantic markings sparingly, as additional colors and superscripts generally clutter the text (see Figure~\ref{fig:phase_marking} and~\ref{fig:level_marking}), although occasionally they were useful tools to explicitly disambiguate parts of a sentence or paragraphs.

We used level markings more often than phase markings.
Most scenarios did not need level markings (11 scenarios), some needed 1-6 level markings (8 scenarios), and one scenario (``UXV Recording'') needed 12 level markings.
In that particular scenario the level markings helped us disentangle a ``ball of yarn'' spun between devices and operating stations.

Phase markings were used scantly.
On the one hand, most scenarios dispensed of phase markings completely (16 scenarios), and very few used them  sparingly (3 scenarios contained one or two phase markings).
On the other hand, one scenario (``Evolving Plan'') specifically benefited from them and used as much as 5 markings to highlight the phases in the text, owning to the scenario scope (plan evolution through different phases).

We thus observe that the markings are an important tool in the toolbox, but one which is to be used selectively.

\subsection{Advantages of BIM as Ubiquitous Language} \label{subsec:advantages_of_bim}

We made a novel choice to use BIM (and its open standard, IFC) as  ubiquitous language (see Section~\ref{subsec:ubiquituous-language}).
This meant that we could often use terms which were well defined from a software engineering perspective, but also well known by many of the AECO professionals as BIM is used widely in construction projects, and a lot of today's AECO software tools and collaboration practices support it.

For example, it might be fuzzy to talk about a ``named 3D area'' on the construction site.
Instead, we used \texttt{IfcZone} which was clearly understood. 
\texttt{IfcZone} is exactly what one can require to be included in the IFC files that will be used as some of the inputs in the software tools that we will develop. 

Furthermore, we chose to use IFC terms for concepts which are not commonly stored in IFC files because we wanted to take advantage of the clear definition existing in the IFC schema.
One example for this is the entity \texttt{IfcTask}.
When we referred to this entity, we knew that it was a task, that it had a start and an end time, that it was part of a workflow and that it could be related to a specific set of building elements. 
Yet another example is \texttt{IfcActor} entertaining well-defined and documented relations such as belonging to an organization or being responsible for a task.

In our 20 scenarios, we specified 98 definitions.
We explicitly did not strive to formalize all of them, and did so only when necessary to avoid ambiguities (see Section~\ref{subsec:ubiquituous-language}).
In total, this gave us only 64 formal definitions (65\%)\footnote{%
The term ``formal'' is used here in a relaxed way.
A formal definition specified properties of an entity, where property types were included in non-obvious cases.}.
Out of these 64 formal definitions, we matched 36 (56\%) to an IFC entity, while the remainder (28, 44\%) were custom data entities specific to our project\footnote{%
We also counted a definition as matching an IFC entity if it specified a super-definition already defined using an IFC entity as in those cases repeating the IFC entity would have been redundant.
}.
In the light of these numbers, we infer that BIM is not a silver bullet, but it provided us with higher precision and easier communication in a substantial fraction of definitions.

\subsection{Effects of the Scenario Space on Scenario Refinement} \label{subsec:effect_of_scenario_space_on_refinement}

During the requirements analysis, we encountered many cases where the requirements were reshuffled, regrouped or abstracted and re-specified, similar to how a programmer refactors the source code.
On one side, some of the refactorings such as concretization of the scenario ``Risk Management'' into ``Fall Risk'' and ``Fire Risk'' (see Figure~\ref{fig:sub-ontology}), were straightforward to derive and we did not  leverage our approach to systematic refinement.
In two notable cases, however, systematic exploration of the scenario space helped us refine the existing requirements in an efficient way so we describe them here as cases in point.

The first case is the dissection of the scenario ``Risk Management'' by the dimension ``phase''.
As the scenario grew larger and unwieldy, we needed to refactored it into smaller parts both for readability and for the ease of future re-writings.
While multiple split axes came into question, using the phase felt the most natural.
Accordingly, the subsequent dissection was fast.
Though it is difficult to compare the refactoring to writing a scenario from scratch, we remark that it took us only 30 minutes to split it up into two (``Fall Risk'' and ``Fire Risk'') and write an overview in a bundling scenario (a new ``Risk Management'').
The split-up can be observed in Figure~\ref{fig:split-by-phase}.

\begin{figure}
    \begin{center}
        \includegraphics[scale=0.45]{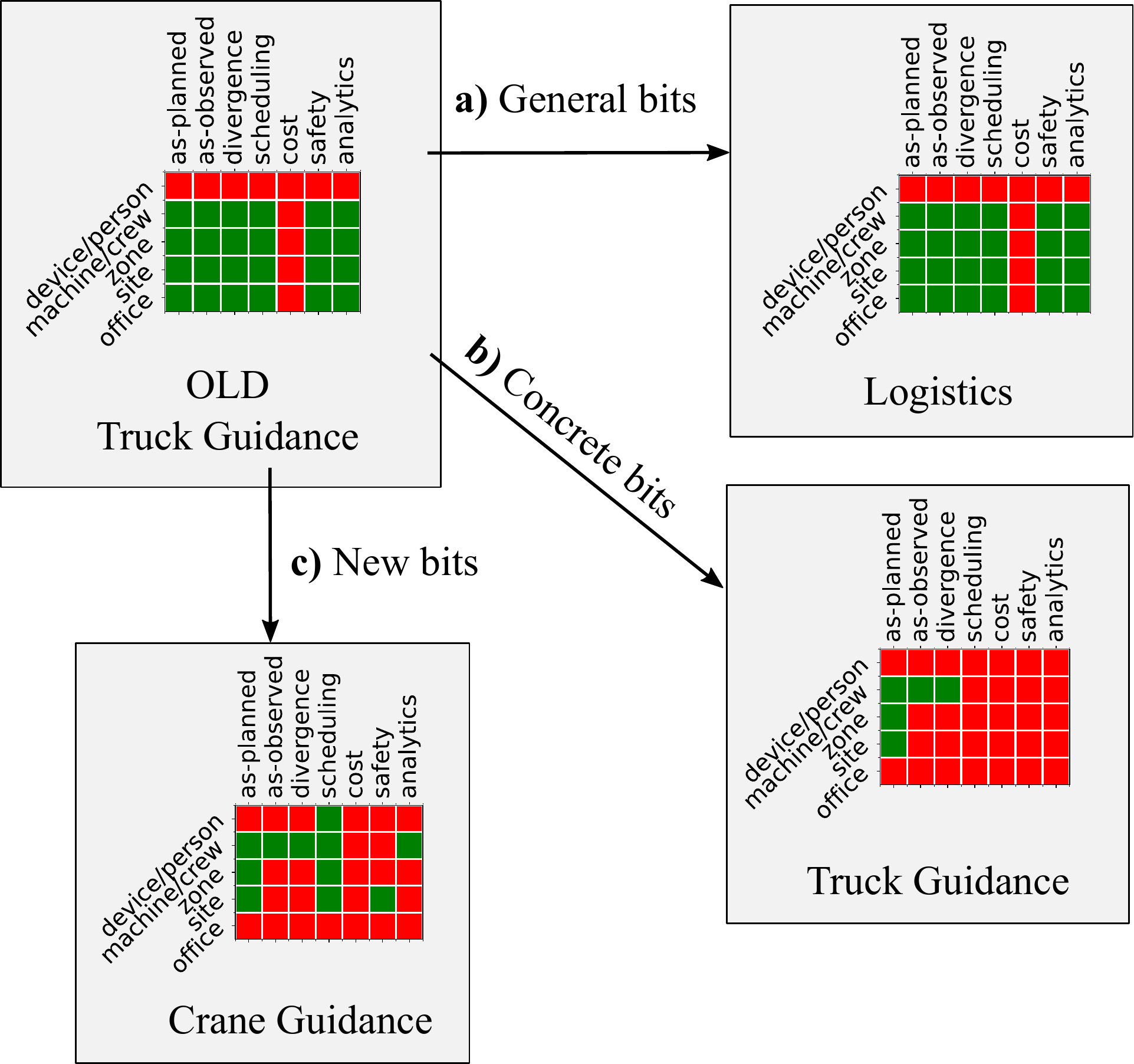}
    \end{center}

    \captionof{figure}{%
    Introduction of a novel scenario, ``Crane Guidance'' through refactoring of the scenario ``Truck Guidance'' from Table~\ref{tab:aspects}a).
    As the scenarios cover only the phase ``construction'', we omit the corresponding dimension for readability.
    \textbf{a)} We extracted the general bits from levels ``site'' and ``office'' from the original scenario ``Truck Guidance'' to the scenario ``Logistics''.
    \textbf{b)} We refactored the details at the level ``machine/crew'' to the new scenario ``Truck Guidance''.
    \textbf{c)} We wrote crane-specific requirements in the new scenario ``Crane Guidance''.
    The markings in the original scenario and the mental framework of the scenario space reduced the effort for introduction of ``Crane Guidance'' to only 1 hour, compared to 2-2.5 hours usually needed for novel scenarios.
    }
    \label{fig:refinement-by-refactoring}
\end{figure}

The second case concerns the introduction of a novel scenario, ``Crane Guidance'' after already writing a scenario on a similar subject, ``Truck Guidance'' (see Table~\ref{tab:aspects}a).
The markings in the text helped us move the  general bits (marked with the level ``site'' and ``office'') to an abstract scenario ``Logistics'', while we kept the truck-specific details (at the level ``machine/crew'') in ``Truck Guidance''.
Afterwards, we added the novel scenario ``Crane Guidance'', where we filled in the specifics for the crane deliveries.
Analogous to the first refinement, we again measured a shorter session until satisfactory results (only 1 hour compared to 2-2.5 hours for a completely novel scenario).
Figure~\ref{fig:refinement-by-refactoring} illustrates the refinement.

While more cases are necessary for a solid evaluation, the two cases presented here suggest that our refinement strategy is beneficial for efficiency and fosters re-use both of concrete requirements and mental efforts. 
In our subjective opinion, we find that our approach strikes a good balance between the rigidness of the structure, compactness, readability, re-usability and modifiability.

\subsection{Generalizability to Other Projects} \label{subsec:generalizability}

Our approach was developed based on the experiences during a single project.
Further studies are necessary to establish its generalizability to other projects.
However, we observe that the conventional approaches (such as~\cite{ bim_governance_bpmn_uml, arayici}, see Section~\ref{sec:related_work} and~\ref{subsec:requirements_analysis}) provided little guidance for the complex nature of the project which surpasses the data governance and ventures into cyber-physical systems.
We anticipate that other AECO projects entailing system interactions with the physical world face similar challenges, which in parallel is also observed in the field of smart manufacturing creating similar solutions~\cite{rami}.

We do not claim the fallibility of the conventional approaches by any means.
They are documented to work well for AECO systems concerning data governance, and it is indeed questionable whether such systems need tools such as a scenario space.
On the other hand, for the development of cyber-physical systems like ours (see Section~\ref{subsec:requirements_analysis}), we believe that we contribute to the literature by documenting our method, our experience using it and its viability.

    \section{Threats to Validity} \label{sec:threats}
Our investigation carries some threats to validity as formulated by~\cite{case_study} regarding the interpretation of the results.

The method proposed here describes how one can capture structured system scenarios in a way that is well suited for complex multi-disciplinary and heterogeneous cases in the AECO industry. 
In the research project~\cite{bimprove} where this method was used, we experienced this as an efficient and well-suited way of capturing system requirements.
But should this method be used by others?
It is very important to consider to what degree our positive experiences of using this method can be used as a evidence for a good fit for other projects.

Regarding \textbf{internal validity} our approach was created as a result of experiences of working with software design and capturing AECO industry needs and improvement potentials over many years. 
However, it is important to notice that this method has not been tried out in external projects. 
For that reason, our paper should be seen as a case study of our experiences. 
Further studies would be needed to know with more certainty if this method works well in other projects.

When it comes to \textbf{external validity}, there was especially one unusual condition of the project. 
We worked with experts from many domains on workflows which were rather innovative for their line of business.
If a project is less innovative, more data-oriented and with little or no interactions with the physical world, the benefits of our approach are questionable as conventional approaches (see Section~\ref{sec:related_work}) are reported to work well.
Our approach is likely to generalize to projects with similar characteristics (\emph{e.g.}, cyber-physical systems of similar scope as described in Section~\ref{subsec:requirements_analysis}).
Whether it generalizes beyond these limits remains an open question.
    \section{Conclusion and Future Work} \label{sec:future_work}

We presented an approach and an accompanying tool for requirements analysis specific to the context of the AECO industry.
The requirements are captured based on scenarios living in a scenario space (along the three dimensions: aspects, phases in a building life cycle and hierarchy levels of details), while the relations between the scenarios are modeled as an ontology.
We developed a command-line program to aid requirements analysis and facilitate the (re-)rendering of the scenario documents.
Finally, we evaluated our approach on 20 practical AECO scenarios from a large AECO project.

In future work, we would like to explore how patterns of common AECO scenarios and relations emerge and how cataloguing them facilitates requirements reuse and accelerates the requirements analysis.
In additional experiments, we would like to validate more thoroughly the adequacy of the scenario space, both its dimensions and its axes.
Finally, a study on a larger scale needs to be performed to assess what individual factors of the approach lead to improvement and degradation of the requirements analysis, respectively, and how our approach quantitatively compares to other methods.

\section*{Acknowledgement}
This work is part of the BIMprove Project, an initiative funded by the European Union’s Horizon 2020 Research and Innovation programme under grant agreement N° 958450, with Focus Area Building a low-carbon, climate resilient future.

    \bibliographystyle{unsrt}
    \bibliography{bibliography}

\begin{thebibliography}{10}

\bibitem{systems_integration_in_architecture}
W.~Shen, Q.~Hao, H.~Mak, J.~Neelamkavil, H.~Xie, J.~Dickinson, R.~Thomas,
  A.~Pardasani, and H.~Xue.
\newblock Systems integration and collaboration in architecture, engineering,
  construction, and facilities management: A review.
\newblock {\em Advanced Engineering Informatics}, 24, 2010.

\bibitem{requirements_engineering_book}
J.~Dick, E.~Hull, and K.~Jackson.
\newblock {\em Requirements Engineering}.
\newblock Springer International Publishing, 2017.

\bibitem{rami}
R.~Heidel, M.~Hoffmeister, M.~Hankel, and U.~Döbrich.
\newblock {\em The Reference Architecture Model {RAMI} 4.0 and the Industrie
  4.0 component}.
\newblock VDE Verlag, 2019.

\bibitem{glinz_scenarios}
Martin Glinz.
\newblock Improving the quality of requirements with scenarios.
\newblock In {\em Second World Congress on Software Quality}, 2000.

\bibitem{IFC_ISO_Standard}
Industry foundation classes ({IFC}) for data sharing in the construction and
  facility management industries — part 1: Data schema.
\newblock Standard ISO 16739-1:2018, International Organization for
  Standardization, 2018.

\bibitem{bim_governance_bpmn_uml}
Eissa Alreshidi, Monjur Mourshed, and Yacine Rezgui.
\newblock Cloud-based {BIM} governance platform requirements and
  specifications: Software engineering approach using {BPMN} and {UML}.
\newblock {\em Journal of Computing in Civil Engineering}, 30, 10 2015.

\bibitem{re_in_aeronautics}
H.~{Gaspard-Boulinc} and S.~{Conversy}.
\newblock Usability insights for requirements engineering tools: A user study
  with practitioners in aeronautics.
\newblock In {\em {IEEE} International Requirements Engineering Conference
  ({RE})}, 2017.

\bibitem{rasaeco_tool}
{RASAECO tool}.
\newblock Retrievable from \url {https://github.com/mristin/rasaeco}.
\newblock Accessed: 2021-04-01.

\bibitem{starchitects}
F.~Cousins.
\newblock Starchitects and jack-hammers: Requirements engineering challenges
  and practices in the construction industry (keynote).
\newblock In {\em {IEEE} International Requirements Engineering Conference
  ({RE})}, 2013.

\bibitem{buildingsmart_use_cases}
{buildingSMART} use case management.
\newblock Retrievable from \url {https://ucm.buildingsmart.org/}.
\newblock Accessed: 2021-04-01.

\bibitem{arayici}
Yusuf Arayici, Vian Ahmed, and Ghassan Aouad.
\newblock A requirements engineering framework for integrated systems
  development for the construction industry.
\newblock {\em Electronic Journal of Information Technology in Construction
  (ITCon)}, 11, 03 2006.

\bibitem{bim_governance}
E.~Alreshidi, M.~Mourshed, and Y.~Rezgui.
\newblock Requirements for cloud-based {BIM} governance solutions to facilitate
  team collaboration in construction projects.
\newblock {\em Requirements Engineering Journal ({REJ})}, 23, 2018.

\bibitem{shafiq}
Muhammad Shafiq, Jane Matthews, and Stephen Lockley.
\newblock A study of {BIM} collaboration requirements and available features in
  existing model collaboration systems.
\newblock {\em Electronic Journal of Information Technology in Construction
  (ITcon)}, 18, 08 2013.

\bibitem{bim_user_requirements}
{C.-E.} Tolmer, C.~Castaing, Y.~Diab, and D.~Morand.
\newblock Adapting {LOD} definition to meet {BIM} uses requirements and data
  modeling for linear infrastructures projects: using system and requirement
  engineering.
\newblock {\em Visualization in Engineering}, 5, 12 2017.

\bibitem{re_ui_in_aeco}
{S.-C.} Chien and A.~Mahdavi.
\newblock Requirement specification and prototyping for user interfaces of
  buildings' environmental controls.
\newblock {\em Electronic Journal of Information Technology in Construction
  (ITcon)}, 14, 2009.

\bibitem{re_for_aeco_simulation_tools}
P.~B. Purup and S.~Petersen.
\newblock Requirement analysis for building performance simulation tools
  conformed to fit design practice.
\newblock {\em Automation in Construction}, 116, 2020.

\bibitem{idm_too_specific}
O.~Berard and J.~Karlshoej.
\newblock Information delivery manuals to integrate building product
  information into desgin.
\newblock {\em Electronic Journal of Information Technology in Construction
  (ITcon)}, 17, 2012.

\bibitem{cps_aeco}
A.~Akanmu, C.~Anumba, and J.~Messner.
\newblock Scenarios for cyber-physical systems integration in construction.
\newblock {\em Electronic Journal of Information Technology in Construction
  (ITCon)}, 18, 2013.

\bibitem{cps_aeco1}
F.~Correa.
\newblock Cyber-physical systems for construction industry.
\newblock In {\em {IEEE} Industrial Cyber-Physical Systems ({ICPS})}, 2018.

\bibitem{requirements_in_the_loop}
W.~Meincke.
\newblock Requirements in the loop - a computer-aided analysis of consistency,
  completeness, and correctness of requirements.
\newblock In {\em {IEEE} International Requirements Engineering Conference
  ({RE})}, 2020.

\bibitem{frame_annotations}
W.~Alhoshan, R.~{Batista-Navarro}, and L.~Zhao.
\newblock Towards a corpus of requirements documents enriched with semantic
  frame annotations.
\newblock In {\em {IEEE} International Requirements Engineering Conference
  ({RE})}, 2018.

\bibitem{requirements_reuse}
X.~Franch, C.~Palomares, and C~Quer.
\newblock Industrial practices on requirements reuse: An interview-based study.
\newblock In {\em Requirements Engineering: Foundation for Software Quality
  ({REFSQ})}, 2020.

\bibitem{guitar_ontology}
H.~Nguyen, J.~Grundy, and M.~Almosy.
\newblock Guitar: An ontology-based automated requirements analysis tool.
\newblock In {\em {IEEE} International Requirements Engineering Conference
  ({RE})}, 2014.

\bibitem{reqpat}
M.~Fockel and J.~Holtmann.
\newblock {ReqPat}: Efficient documentation of high-quality requirements using
  controlled natural language.
\newblock In {\em {IEEE} International Requirements Engineering Conference
  ({RE})}, 2015.

\bibitem{visualization_review}
Zahra Shakeri, Mohammad Noaeen, and Guenther Ruhe.
\newblock Requirements engineering visualization: A systematic literature
  review.
\newblock In {\em {IEEE} International Requirements Engineering Conference
  ({RE})}, 2016.

\bibitem{shapeRE}
Y.~D. Pham, A.~Bouraffa, and W.~Maalej.
\newblock Shapere: Towards a multi-dimensional representation for requirements
  of sustainable software.
\newblock In {\em {IEEE} International Requirements Engineering Conference
  ({RE})}, 2020.

\bibitem{bimprove}
{BIMprove Project}.
\newblock Retrievable from \url {https://www.bimprove-h2020.eu}.
\newblock Accessed: 2021-04-01.

\bibitem{dronology}
J.~{Cleland-Huang} and M.~{Vierhauser}.
\newblock Discovering, analyzing, and managing safety stories in agile
  projects.
\newblock In {\em {IEEE} International Requirements Engineering Conference
  ({RE})}, 2018.

\bibitem{novices_in_requirements}
M.~Bano, D.~Zowghi, A.~Ferrari, P.~Sploetini, and B.~Donati.
\newblock Learning from mistakes: An empirical study of elicitation interviews
  performed by novices.
\newblock In {\em {IEEE} International Requirements Engineering Conference
  ({RE})}, 2018.

\bibitem{bim_dimensionality}
A.~Koutamanis.
\newblock Dimensionality in {BIM}: Why {BIM} cannot have more than four
  dimensions?
\newblock {\em Automation in Construction}, 114, 2020.

\bibitem{bimprove_scenarios}
{BIMprove Scenario Ontology}.
\newblock Retrievable from \url
  {https://mristin.github.com/bimprove-scenarios}.
\newblock Accessed: 2021-04-01.

\bibitem{riba}
{RIBA} plan of work.
\newblock Retrievable from \url
  {https://www.architecture.com/knowledge-and-resources/resources-landing-page/riba-plan-of-work}.
\newblock Accessed: 2021-04-01.

\bibitem{requirements_zoom}
P.~Ghazi and M.~Glinz.
\newblock An experimental comparison of two navigation techniques for
  requirements modeling tools.
\newblock In {\em {IEEE} International Requirements Engineering Conference
  ({RE})}, 2018.

\bibitem{domain_knowledge_important}
A.~M. Aranda, O.~Dieste, and N.~Juristo.
\newblock Effect of domain knowledge on elicitation effectiveness: An
  internally replicated controlled experiment.
\newblock {\em {IEEE} Transactions on Software Engineering}, 42, 2016.

\bibitem{domain_drive_design}
E.~J. Evans.
\newblock {\em Domain-Driven Design: Tackling Complexity in the Heart of
  Software}.
\newblock Addison Wesley Longman, 2003.

\bibitem{uml_too_heavy_weight}
M.~Glinz.
\newblock Problems and deficiencies of {UML} as a requirements specification
  language.
\newblock In {\em International Workshop on Software Specification and Design},
  2000.

\bibitem{bim_in_peru}
C.~{Sanchís-Pedregosa}, J.~{Vizcarra Aparicio}, and A.~{Leal-Rodríguez}.
\newblock {BIM}: a technology acceptance model in {Peru}.
\newblock {\em Electronic Journal of Information Technology in Construction
  (ITCon)}, 25, 2020.

\bibitem{bim_in_malaysia}
W.~Enegbuma, G.~Aliagha, and K.~Ali.
\newblock Preliminary building information modelling adoption model in
  {Malaysia}.
\newblock {\em Construction Innovation}, 14, 2014.

\bibitem{listens_learned}
Alistair Mavin, Philip Wilkinson, Sarah Gregory, and Eero Uusitalo.
\newblock Listens learned (8 lessons learned applying {EARS} ).
\newblock In {\em {IEEE} International Requirements Engineering Conference
  ({RE})}, 2016.

\bibitem{comprehensible_visual_notation}
P.~Caire, N.~Genon, P.~Heymans, and D.~L. Moody.
\newblock Visual notation design 2.0: Towards user comprehensible requirements
  engineering notations.
\newblock In {\em {IEEE} International Requirements Engineering Conference
  ({RE})}, 2013.

\bibitem{sensemaking}
P.~Ralph and R.~Mohanani.
\newblock Is requirements engineering inherently counterproductive?
\newblock In {\em IEEE/ACM International Workshop on the Twin Peaks of
  Requirements and Architecture ({TwinPeaks})}, 2015.

\bibitem{glinz_statecharts}
M.~Glinz.
\newblock An integrated formal model of scenarios based on statecharts.
\newblock In {\em European Software Engineering Conference ({ESEC})}, 1995.

\bibitem{lightweight_requirements_assesments}
D.~Rapp, A.~Hess, N.~Seyff, P.~Spörri, E.~Fuchs, and M.~Glinz.
\newblock Lightweight requirements engineering assessments in software
  projects.
\newblock In {\em {IEEE} International Requirements Engineering Conference
  ({RE})}, 2014.

\bibitem{bimprove_scenarios_src}
{BIMprove Scenario Ontology (source code)}.
\newblock Retrievable from \url
  {https://github.com/mristin/bimprove-scenarios}.
\newblock Accessed: 2021-04-01.

\bibitem{quality_user_stories}
G.~Lucassen, F.~Dalpiaz, J.~v.~d. Werf, and S.~Brinkkemper.
\newblock Improving agile requirements: the quality user story framework and
  tool.
\newblock In {\em {IEEE} International Requirements Engineering Conference
  ({RE})}, 2016.

\bibitem{markdown}
S.~Leonard.
\newblock Guidance on markdown: Design philosophies, stability strategies, and
  select registrations.
\newblock Technical report, Internet Engineering Task Force ({IETF}), 2016.

\bibitem{pro_git}
S.~Chacon and B.~Straub.
\newblock {\em Pro Git}.
\newblock Apress, 2014.

\bibitem{case_study}
Robert~K. Yin.
\newblock {\em Case study research and applications: Design and methods}.
\newblock Sage publications, 2017.

\end{thebibliography}
\end{document}